\documentclass[10pt,twocolumn,twoside]{IEEEtran}
\IEEEoverridecommandlockouts
\usepackage{cite}
\usepackage{amsmath,amssymb,amsfonts}
\usepackage{algorithmic}
\usepackage{graphicx}
\usepackage{textcomp}
\usepackage{xcolor}
\usepackage{algorithm}
\usepackage{comment}
\usepackage{setspace}
\usepackage{color}
\usepackage{booktabs}
\usepackage{multirow}
\usepackage{subfigure}

\def\BibTeX{{\rm B\kern-.05em{\sc i\kern-.025em b}\kern-.08em
    T\kern-.1667em\lower.7ex\hbox{E}\kern-.125emX}}
\begin{document}
\title{Deep Reinforcement Learning for Fog Computing-based Vehicular System with Multi-operator Support\\
}
\author{\author{\IEEEauthorblockN{Xiaohan~Zhang\IEEEauthorrefmark{1}, Yong~Xiao\IEEEauthorrefmark{1}, Qiang Li\IEEEauthorrefmark{1}, Walid Saad \IEEEauthorrefmark{2}\\
\IEEEauthorblockA{\IEEEauthorrefmark{1}School of Electronic Information and Communications, Huazhong Univ. of Science \& Technology, China}\\
\IEEEauthorblockA{\IEEEauthorrefmark{2} Department of Electrical and Computer Engineering, Virginia Tech, VA}
}}}
\maketitle

\begin{abstract}
This paper studies the potential performance improvement that can be achieved by enabling multi-operator wireless connectivity for cloud/fog computing-connected vehicular systems. Mobile network operator (MNO) selection and switching problem is formulated by jointly considering switching cost, quality-of-service (QoS) variations between MNOs, and the different prices that can be charged by different MNOs as well as cloud and fog servers. A double deep Q network (DQN) based switching policy is proposed and proved to be able to minimize the long-term average cost of each vehicle with guaranteed latency and reliability performance. The performance of the proposed approach is evaluated using the dataset collected in a commercially available city-wide LTE network. Simulation results show that our proposed policy can significantly reduce the cost paid by each fog/cloud-connected vehicle with guaranteed latency services.
\end{abstract}

\section{Introduction}
Cloud and fog computing-based vehicular systems have recently been  embraced by both industry and standardization institutions as a promising solution to support computationally intensive services with low-latency requirements by deploying a large number of low-cost fog nodes close to the users.  
One of the key pre-requisites for the success of such a system is to maintain a ubiquitously available wireless connectivity with ultra-low latency and ultra high reliability data transferring links between vehicles and cloud data center and fog nodes. Unfortunately, due to the random nature of wireless channel, it is generally impossible to always maintain a performance-guaranteed wireless link when driving into different locations across a wide-geographical area\cite{6064713,7399689}. One possible solution is to adopt a multi-operator approach. Recent results\cite{xiao2019driving} as well as our own measurement suggest that the performance of different mobile network operators (MNOs) may exhibit strong spatial and temporal variations. By allowing a connected-vehicle to dynamically switch to the MNO's network with the lowest latency and the highest reliability, it has the potential to significantly improve the performance of the wireless connectivity as well as the service coverage for cloud and fog-supported vehicular system without deploying new wireless network infrastructure. 



Despite of its great promise, multi-operator-supported vehicular system has been hindered by the following challenges:
 \begin{itemize}
 \item[1)] The performance of wireless connectivity offered by different MNOs is typically temporally-spatially varying, unpredictable and uncontrollable. It is generally impossible for each vehicle to instantaneously detect the best-performed MNO's networks and make the switching decision.
 \item[2)] Frequent switching between the networks of MNOs can result in extra cost, communication overhead as well as computational load. Also, switching between different MNO's networks and base stations (BSs) may result in increased latency and unreliability of wireless connection.
 \item[3)] High speed vehicles may move into different locations from time to time. Keeping track of the changes for the fast-changing networking environment is known to be a notoriously challenging problem.
\end{itemize}

The main contribution of this paper is to address the above challenges by proposing a novel multi-operator switching policy for each connected vehicle to dynamically switch between different MNOs' networks as well as the connected cloud and fog servers. As such, the service latency and reliability performance can be guaranteed throughout the entire driving route with minimized cost for switching among different MNOs' networks. We summarize our main contributions as follows:
\begin{itemize}
\item[1)] We formulate the multi-operator switching problem for a cloud and fog-supported vehicular system as a dynamic programming problem by taking into consideration of the MNO switching cost, service variations of connected vehicle, and different price that can be charged by cloud and fog servers when connecting to different MNOs.
\item[2)] We derive the optimal volume of workload allocated to fog and cloud servers when connecting to each MNO's network and then propose a double deep Q networks (DQN) based MNO switching policy to minimize the overall cost paid by each vehicle with guaranteed latency and reliability performance. We prove that our proposed policy can always approach the optimal MNO switching policy with much less computational complexity compared to the traditional Q-learning based approach.
\item[3)] We evaluate the performance of our proposed approach by using the round-trip-time (RTT) datasets collected in a commercially available LTE network for over four months of measurement throughout a mid-sized city. Simulation results show that our proposed policy can significantly reduce the cost paid by each fog/cloud-supported vehicle with guaranteed latency services when driving into different locations.
\end{itemize}

\section{Related Works}
Fog computing has been considered as one of the key solutions to support computational-intensive services with  low end-to-end latency requirements\cite{xiao2018fog,deng2016optimal,xiao2018,lee2019online}. However, due to the limit in cost, space, and resource, the computing capability of each individual fog node is quite limited, especially compared with the cloud data center. Many existing works focused on balancing the workload between fog nodes and cloud data center. 
In \cite{deng2016optimal}, the authors proposed an optimal workload allocation scheme to minimize the power consumption of fog nodes with constraints on service delay. An online, low-complexity algorithm for joint resource allocation among collaborative fog nodes was proposed in \cite{lee2019online}.

Fog computing has also been applied in smart vehicular networks to meet the stringent demand on latency-sensitive services such as autonomous driving and intelligent driving assistance\cite{xiao2018,5GAA,zhang2017mobile,zeng2019joint}. In \cite{zhang2017mobile}, the authors proposed a novel offloading framework in smart vehicle systems to reduce the end-to-end latency and transmission cost. 
In \cite{zeng2019joint}, a novel framework was introduced to optimize the operation of vehicle platoon with constraints on the delay of the wireless vehicle-to-vehicle (V2V) networks and the stability of the vehicular system.

Multi-operator resource sharing has attracted significant interest recently due to its capability of expending the accessible resource of each MNO without the need to deploy any new infrastructure. Most existing works focused on jointly optimizing the resource utilization among MNOs. For instance, the authors in \cite{7124517} introduced an infrastructure-sharing approach for MNOs to collaborate with each other to reduce the energy consumption by deactivating the under-utilized base stations. In \cite{sugathapala2015quantifying}, the authors 
investigated various benefits generated by spectrum sharing for all MNOs when multi-operator cooperation can be enabled.

\section{Background and Architecture Overview}
We consider a cloud and fog supported vehicular system consisting of a set $\mathcal{M}$ of $M$ MNOs that can offer wireless connection between vehicles and servers at the cloud data centers (CDCs) or fog nodes. We assume that each MNO has deployed an exclusive set of fog nodes inside of its wireless network infrastructure. Let $\mathcal{F}_{m}$ be the set of fog nodes in MNO $m$'s network, for $m\in\mathcal{M}$. Each connected vehicle can generate a sequence of workloads to be submitted and processed by servers at fog nodes and CDC.

Without loss of generality, we consider slot-based process and assume each vehicle can only connect to one MNO's network during each time slot. Let $m_{t}$ be the MNO selected by the considered vehicle in time slot $t$ for $m_{t} \in \mathcal{M}$. We follow a commonly adopted setting and assume the workload generation process of each vehicle in each time slot follows a Poisson distribution with parameter $\lambda_{t}$, i.e., we write the number of workloads $w_{t}$ generated by each vehicle during time slot $t$ as $w_{t} \sim \mathcal{P}(\lambda_{t})$.

At the beginning of each time slot, each vehicle can decide whether or not to switch to another MNO's network or maintain its current choice. Once a vehicle makes its decision, it cannot switch to another MNO's network during the rest of the time slot. We consider a dynamic decision problem with a finite horizon, in which a vehicle driving from the start point to the final destination experiences $T$ time slots of driving. The vehicle can dynamically switch between different MNOs' networks across different time slots with the goal of optimizing its average service performance during the entire route. Note that switching between different MNOs' networks generally results in extra delay that can be caused by service handover and resource release and re-allocation. To simplify our discussion, we assume the extra delay for each vehicle to switch between any two MNOs' networks to be a constant, denoted as $d$ ms.
\begin{table}[t] 
\centering
\caption{Service level latency and reliability  illustrated in\cite{5GAAUseCases}}
\vspace{-0.1in}
\begin{tabular} {p{4.25cm}p{1.5cm}p{1.5cm}}\toprule
\hline
Type of services &  Service Level Latency($\tau$)  &  Service Level Reliability($\gamma$)\\ 
\hline 
Cross-Traffic Left-Turn Assist & $100$ ms & $90\%$  \\
\hline
Emergency Brake Warning & $120$ ms& $99.9\%$  \\
\hline
Lane Change Warning & $400$ ms& $99.9\%$  \\
\hline
\end{tabular}
\label{table_service}
\vspace{-0.2in}
\end{table}

As observed in \cite{5GAAUseCases}, a connected vehicle can request different services when driving into different locations. In this paper, we use $\mathcal{X}$ to denote the set of all possible service. For example, a vehicle trying to take left-turn at a cross-road location will be more likely to  request accurate driving trajectory guidance service. When driving on a highway, however, a vehicle is more likely to periodically request route information update and front road condition report. In general, different services may have different latency and reliability requirements. We have listed several typical connected vehicular services as well as their corresponding latency and reliability requirements reported in \cite{5GAAUseCases} in Table \ref{table_service}.
In our previous work \cite{xiao2019driving}, we have already observed that service latency characterized by the RTT between a vehicle and a given server at a fog node or CDC at given location can be considered as a stationary probability distribution. Let $\Pr^{c}(\eta,l_t,m_t)$ and $\Pr^{f}(\eta,l_{t},m_t)$ be the probability distribution functions of service latency $\eta$ between vehicle and CDC and that between vehicle and the closest fog node when connecting to MNO $m_{t}$ at location $l_t$ \footnote{Note that $\Pr^{c}(\eta,l_{t},m_t)$ and $\Pr^{f}(\eta,l_{t},m_t)$ are also related to the driving speed. However, it has been observed in \cite{xiao2019driving} that the driving speed of the vehicle at a given location falls into a specific range. We therefore let  $\Pr^{c}(\eta,l_{t},m_t)$ and $\Pr^{f}(\eta,l_{t},m_t)$ are the average probability that the vehicle driving into a given location $l_t$ within the given range of driving speed.}. Each service  $x$ requested by the vehicle at location $l_t$ has a specific requirement characterized by the service confidence level, which is defined as the probability that a given service latency $\tau_x$ ms can be satisfied, i.e., $f(x,m_t,l_t)=\Pr(\eta \leq \tau_x \mid l_t,m_t)=\int_{\eta=0}^{\tau_x} \Pr(\eta,l_t,m_t)\, d \eta$ where $\Pr(\eta,l_t,m_t)$ is the probability that the service latency in the location $l_t$ when connecting to MNO $m_t$ is $\eta$ ms. Note that if the vehicle changes the network at time slot t, the service level latency $\tau_x$ corresponding to the generated service $x$ needs to be reduced to overcome the extra switching delay. We assume there is a lower bound $\gamma_x$ on the confidence level for each service requested at location $l_t$, i.e., $f(x,m_t,l_t) \geq \gamma_x$. For example, as shown in Table \ref{table_service}, the requirement of cross-traffic left-turn assistant service can be written as $f(\tau=100 $ ms$,m_t,l_t) \geq 0.9$.

In our model, we assume that all MNOs have service coverage throughout the entire driving route. Fog nodes and CDC charge different prices for workload processing service. Let $\mu$ \$ per TB and $\nu$ \$ per TB be the prices charged by fog nodes and CDC for processing each workload respectively. In general, we have
\begin{eqnarray}
\mu > \nu.
\label{condition}
\end{eqnarray}

In other words, if the CDC can offer service with satisfactory latency requirement, each vehicle should always submit all of their workload to the CDC. In here, we consider a more general setting and assume each vehicle in each time slot $t$ can offload its workload to the fog node of the selected MNO with probability $\alpha_t$ and submit the workload to the CDC with probability ($1-\alpha_t$), i.e., we write the total cost paid by the vehicle in time slot $t$ as
\begin{eqnarray}
\varpi(m_t,\alpha_t) = \mu \alpha_t w_t + \nu (1-\alpha_t) w_t.
\end{eqnarray}

It is therefore very important for the vehicle to choose the MNO's network that offers satisfactory service performance with the lowest cost.

In other words, we can write the optimization problem for the considered vehicle as follows:
\begin{eqnarray}
\begin{split}
\min\limits_{\boldsymbol{\alpha},\boldsymbol{m}}\  \mathbb{E} \{&\sum\limits_{t=1}^{T} \varpi(m_t,\alpha_t)\}\\
{\mbox{s.t.}}\  &f(x_t,m_t,l_t) \geq \gamma_{x_t},
\end{split}
\label{object}
\end{eqnarray}
where $\boldsymbol{\alpha}$ and $\boldsymbol{m}$ denote the vectors of all parameters $\alpha_t$ and $m_t$ at all time slots respectively, i.e., $\boldsymbol{\alpha}=\langle \alpha_t \rangle,\boldsymbol{m}=\langle m_t \rangle,\forall t \in \{1,2,3...,T\}$.

\textbf{Remark:}  We assume in each location $l_t$ on the driving path, there is always at least one MNO that can offer satisfactory service to the vehicle, i.e., we have $\exists m_t \in \mathcal{M},f(x_t,m_t,l_t) \geq \gamma_{x_t},\forall l_t \in \mathcal{L}$, where $\mathcal{L}$ is the set of all possible locations of the driving route. Note that the service confidence level offered by different MNOs can vary from one time slot to another. However, it is generally impractical for each vehicle to always switch to the MNO that offers the highest service confidence level at each time slot, especially considering the extra delay caused by MNO switching. Each vehicle must carefully decide a sequence of MNOs with minimized number of MNO switching, so the average cost can be minimized while the required confidence level can be satisfied throughout the entire journey.

\section{Optimization for MNO Switching}

We observe that solving problem (\ref{object}) requires each vehicle to careful select a sequence of MNO networks and the connected fog nodes and cloud data center  when driving into each location. 
In particular, when deciding the optimal MNO to connect, each vehicle not only needs to evaluate the network performance of different MNOs but also the optimal amount of workload that can be processed by the corresponding fog node and cloud data center. Each vehicle also needs to evaluate the cost for MNO switching, i.e., the vehicle should not switch to another MNO if the MNO switching cost exceeds the current and future performance benefit achieved by the switching. 

\subsection{\emph{Optimal Task Assignment} }
Let us first consider the case that a vehicle has already selected an MNO at a given location in time slot $t$. This vehicle will then need to carefully decide the portion of the workload to be allocated to the fog node and the cloud data center. More specifically, the vehicle needs to solve the following optimization problem,
\begin{eqnarray}
\begin{split}
{\mathop{\min}\limits_{\alpha_t}} \ & \mu \alpha_t w_t+\nu (1-\alpha_t)  w_t      \\
{\mbox{s.t.}}\  &f(x_t,m_t,l_t) \geq \gamma_{x_t},\\
& 0\leq \alpha_t \leq 1.
\end{split}
\label{alpha}
\end{eqnarray}

According to (\ref{condition}), we can rewrite the objected as follows:
\begin{eqnarray}
\begin{split}
\mathop{\min}\limits_{\alpha_t}\  & \alpha_t      \\
\mbox{s.t.}\ &f^{c}(x_t,m_t,l_t){(1-\alpha_t)}+f^f(x_t,m_t,l_t){\alpha_t} \geq \gamma_{x_t},\\
& 0\leq \alpha_{t} \leq 1.
\end{split}
\label{alpha}
\end{eqnarray}

Hence, the optimal task assignment is
\begin{eqnarray}
\alpha^*_t(m_t,l_t,x_t) = {({\gamma_{x_t}-{f^{c}(x_t,m_t,l_t)}})\over {(f^{f}(x_t,m_t,l_t)-f^{c}(x_t,m_t,l_t))}}.
\end{eqnarray}
\subsection{\emph{Finite Horizon MNO Switching Problem}}
In the MNO switching problem, we need to carefully decide which MNO we should select or switch to at different time slots to maximize the long-term performance. Suppose that the state in this time slot only depends on the state and switching decision made in the previous time slot. We can then formulate the network switching problem as a Markov decision Process (MDP) with finite horizon consisting of the following components:

$\bullet$ \emph{State space} $\mathcal{S}$: is a finite set of the vehicle's location $l$, the generated service $x$, and the selected MNO $m$. We write the state in time slot $t$ as $s_t= \langle l_t,x_t,m_t \rangle$ for $l_t \in{\mathcal{L}}$, $x_t \in \mathcal{X}$, and $m_t \in \mathcal{M}$.

$\bullet$ \emph{Action space} $\mathcal{A}$: is a finite set of all possible MNOs which can be connected to. We write an instance of action at time slot $t$ as $a_t$ for $a_t \in{\mathcal{A}}$.

$\bullet$ \emph{State transition function} $\mathcal{T}$: $\mathcal{S}\times\mathcal{A}\times\mathcal{S}\longrightarrow[0,1]$: the probability of state transiting from one state to another. In particular, the probability of the state transferred from state $s$ to $s'$ when taking action $a$ can be expressed as $\mathcal{T}(s',s,a)=\Pr(s'|s,a)$. In this paper, the type of service and the location only depends on the service type and the location in the previous time slot,
Without loss of generality, we assume the transition probability of location and requested service is independent with the selection of MNOs, i.e., we have $\Pr(s'|s,a)=\Pr(x',l'|x,l)\Pr(m'|m,a)$. This assumption is reasonable because in most practical scenarios, the local and service requested by users are depending on the users' preference and driving behavior both of which are typically independent with the MNO connectivity.  To simplify the problem, we assume $\Pr(x',l'|x,l)$ can be pre-calculated and assumed to be known by the vehicle.
The MNO $m'$ selected in the next time slot will be decided by the MNO selected action $a$. We assume the vehicle can always connect to the intended MNO according to its MNO selection, i.e., we assume $\Pr(m'|m,a)=1$ if $a=m'$.

$\bullet$ \emph{Utility function}: The main objective is to minimize the total cost by choosing the optimal action at each time slot. We follow a commonly adopted setting and write the utility function as $R_{t}(s_t,a_t)=\varpi(m_t,\alpha_t^{*}(s_t))$ at state $s_t$ and time slot $t$. An MNO switching policy is defined below:

\textbf{Definition 1: switching policy} is a function mapping the time slot and the corresponding state $s$ into an action:
\begin{eqnarray}
\pi:\{1,...,T\} \times {\cal S} \rightarrow {\cal A}.
\end{eqnarray}

The total expected utility achieved by policy $\pi$ for initial state $s_0$ is given by
\begin{eqnarray}
V^{\pi}=\mathbb{E}\{\sum\limits_{t=1}^{T} R_{t}(s_t,a_t)|s_0,\pi\}.
\end{eqnarray}
\subsection{Double DQN approach for MNO switching}

\subsubsection{Q-learning method}
As a reinforcement learning method, Q-learning\cite{watkins1992q} is usually chosen to find the optimal policy for sequential decision problems. In particular, the Q-learning algorithm can be implemented based on a value table commonly referred to as Q-table which is used for storing all the possible Q-value, the expected long-term utility that can be achieved by various possible pairs of state and action. The vehicle will choose an action that minimize the Q-value when a state has been observed.  
Q-learning algorithm will also update the Q-value according to the observed results, which consist of the current utility and the state in the next time slot. This process will be repeated. Q-learning algorithm can always learn from the previous decisions and adjust its Q-table and the corresponding policy accordingly.  
It has already been proved that Q-learning can always converge to an optimal policy after a finite number of iterations\cite{watkins1992q}. In this paper, our main objective is to obtain the optimal switching policy $\pi^*: \mathcal{S} \times \{1,2,...,T\} \rightarrow \mathcal{A}$ for the vehicle to minimize its long-term average cost.


The optimal 
policy can be written as follows. 
\begin{eqnarray}
a^*_t=\arg \min\limits_{a_t\in \mathcal{A}}\ {\mathcal{Q}(s_t,a_t)},
\end{eqnarray}
where
the Q-function is evolved as follows: 
\begin{eqnarray}
\begin{split}
\mathcal{Q}(s_t,a_t)&=\mathcal{Q}(s_t,a_t)+g[R_t(s_t,a_t)\\
&+b\min\limits_{a_{t+1}}\ \mathcal{Q}(s_{t+1},a_{t+1})-\mathcal{Q}(s_t,a_t)].
\label{Q-value}
\end{split}
\end{eqnarray}
where $0 \leq b \leq 1$ denotes the discount factor which embodies the weight of the long-term reward \cite{watkins1992q}, and $g$ is the learning rate which represents the influence of the new value to the existing one. 

\subsubsection{Double deep Q network}

\begin{algorithm}
\setstretch{1.2}
\footnotesize
  \caption{MNO Switching Policy Based on Double DQN}\label{Algorithm 2}
  \begin{itemize}
    \item[\textbf{1)}] Initialize the \emph{replay memory pool}
    \item[\textbf{2)}] Initialize the Q-network $\mathcal{Q}$ and the target Q-network $\hat{\mathcal{Q}}$ with arbitrary weight
    $\boldsymbol{\vartheta}$ and  $\boldsymbol{\vartheta}^-$
    \item[] {\bf For} $episode=1\ to\ N$ {\bf do}
          \begin{itemize}
            \item[\textbf{1)}]Set $t=1$ and observe the initial state $s_1$
            \item[]{\bf Repeat:}
                \begin{itemize}
                    \item[\textbf{1)}] Select an arbitrary action $a_{t}$ with probability $\epsilon$ or a deliberate action $a_{t}= \arg\min\limits_{a_{t}}\mathcal{Q}(s_t,a_{t};\boldsymbol{\vartheta)}$ with probability $1-\epsilon$
                    \item[\textbf{2)}] Obtain the observation including the immediate reward $R_t$ and the next state $s_{t+1}$ and putting the transition $(s_t,a_t,R_t,s_{t+1})$ into the memory pool.
                    \item[\textbf{3)}] $t=t+1$
                \end{itemize}
            \item[] {\bf Until} $s_{t+1}$ is the terminal state.
            \item[\textbf{2)}]Sample minibatch of experience $(s_i,a_i,R_i,s_{i+1})$ from the memory pool arbitrarily.
            \item[\textbf{3)}]Compute $y_i=R_i+g\mathcal{Q}(s_{i+1},\arg\min\limits_{a_{i+1}}\hat{\mathcal{Q}}(s_{i+1},a_{i+1};\boldsymbol{\vartheta});\boldsymbol{\vartheta}^-)$
            \item[\textbf{4)}]Use the gradient descent method on $(y_i-\mathcal{Q}(s_i,a_i;\boldsymbol{\vartheta}))^2$  with regard to the parameter $\boldsymbol{\vartheta}$
            \item[\textbf{5)}] Update the $\hat{\mathcal{Q}}=\mathcal{Q}$ for every $\mathcal{C}$ steps
    \end{itemize}
	\item[] {\bf end for}
  \end{itemize}
\end{algorithm}
\begin{figure}[!htb]
  \centering
  \vspace{-0.1in}
  \includegraphics[width=6cm]{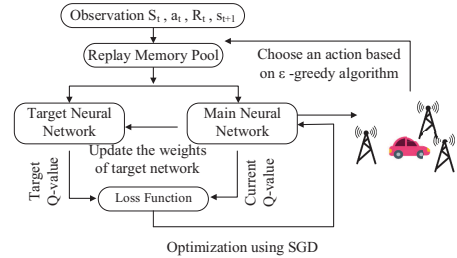}
  \vspace{-0.1in}
  \caption{The proposed double DQN-based framework. }
  \label{deepfig}
\end{figure}

Since Q-table needs to store all the possible pairs of state and action in each time slot, a large amount of storage space is required and will typically result in a slow convergence rate. 
The double DQN algorithm\cite{van2016deep} is introduced to address the above problem. The key idea of the double DQN algorithm is to introduce a deep neural network, referred to as the primary Q-network, to select an action. To stabilize the primary Q-networks, another deep neural network, called the target Q-network, is introduced to frequently (but slowly) update to the primary Q-network values to reduce the correlations between the target and estimated Q-values, thereby stabilizing the algorithm.

We describe the architecture of our proposed algorithm in Fig \ref{deepfig}. Specifically, the training process consists of many episodes. 
In each episode, an action will be selected based on the $\epsilon$-greedy algorithm. In this algorithm, the vehicle will randomly choose an arbitrary action with probability $\epsilon$ in each episode. Otherwise, an  optimal  action which minimizes the Q-value  $\mathcal{Q}(s_t,a_{t};\boldsymbol{\vartheta})$ will be selected. In addition, the algorithm initializes parameter $\epsilon$ with a large value, e.g., 0.8, in the beginning and gradually (e.g., linear rate or exponential rate) decreases the parameter to a small value such as 0.05. In this way, the action of the vehicle will be randomly selected at the beginning and will then gradually approach to a deterministic policy with high probability to make the optimal decision. For instance, the vehicle will choose an action $a_t$ for the current state $s_t$ and obtain the current utility $R_t$ and 
observe the next state $s_{t+1}$. The observation $(s_t,a_t,R_t,s_{t+1})$ will then be put into a memory, called replay memory pool\cite{fang2017learning}. 

After that, the learning process will be executed based on random samples from the pool. In this way, the data would be more likely to be independent and the transitions generated at the previous episodes can be exploited more times.
The main role of the training process is to use the deep neural network to estimate the Q-value for each action in a given state at each time slot. The neural network is trained based on the random samples of the previous experience stored in the replay memory pool. Specifically, we consider three main features, location, type of service, and the selected MNO, for each state at each time slot. 

We then use the \emph{stochastic gradient descent} algorithm to optimize the primary Q-network's parameter, denoted as $\boldsymbol{\vartheta}$.  
The basic idea of using stochastic gradient descent is to estimate the gradient based on a small set of samples. Generally, a minibatch of experience will be sampled uniformly from the memory pool during each episode and the minibatch size denoted as $D'$ is set to be  a small number compared with the number of experiences in the pool. The estimation of the gradient can be calculated as 
\begin{eqnarray}
\begin{split}
 \boldsymbol{\zeta} = {1 \over D'}\nabla_{\boldsymbol{\vartheta}}\sum\limits_{i=1}^{D'}L((s,a,R,s')^{(i)},\boldsymbol{\vartheta}).
\end{split}
\end{eqnarray}
where $L((s,a,R,s^{'})^{(i)},\boldsymbol{\vartheta})$ is the loss function defined as
\begin{eqnarray}
\lefteqn{L((s,a,R,s')^{(i)},\boldsymbol{\vartheta})} \nonumber \\
&=&\mathbb{E}_{(s,a,R,s')}[((R+g\mathcal{Q}(s',\arg\min\limits_{a'}\hat{\mathcal{Q}}(s',a';\boldsymbol{\vartheta}_i); \nonumber \\
&&\boldsymbol{\vartheta}_i^{-}) -\mathcal{Q}(s,a;\boldsymbol{\vartheta}_i))^2],
\label{loss function}
\end{eqnarray}
where $\boldsymbol{\vartheta}_i$ and $\boldsymbol{\vartheta}_i^{-}$ denote the parameter of the primary Q-network and the target Q-network, respectively.

The stochastic gradient decent algorithm can then update the parameter as
\begin{eqnarray}
\begin{split}
 \boldsymbol{\vartheta}\leftarrow\boldsymbol{\vartheta}-\upsilon\boldsymbol{\zeta}\
\end{split}
\label{downhill}
\end{eqnarray}
where $\upsilon$ is the learning rate.

Note that the algorithm only replaces the parameter $\boldsymbol{\vartheta}_i^{-}$ of the target neural network by the Q-network parameter $\boldsymbol{\vartheta}_i$ every $\mathcal{C}$ episodes. Hence, the main Q-network value would be updated more smoothly by the target Q-network every $\mathcal{C}$ episodes. More details of our proposed algorithm  is described in \textbf{Algorithm \ref{Algorithm 2}}.

\section{Experimental Results}
	\begin{figure*}[!htbp]
	\subfigure[]{
	\begin{minipage}[t]{0.23\linewidth}
	\centering
	\includegraphics[width=4cm]{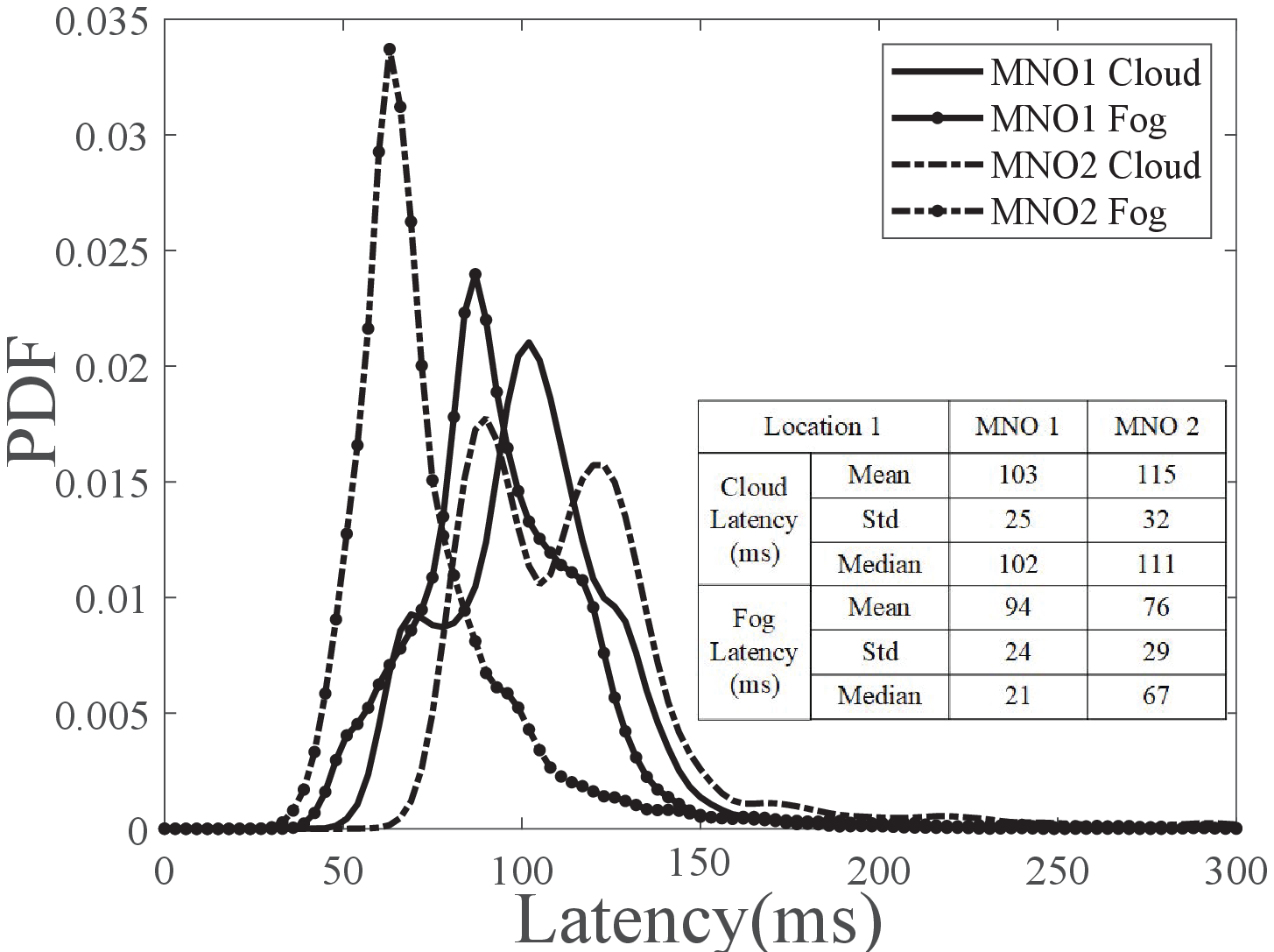}
	\end{minipage}%
    \vspace{-2in}
	\label{MNO1perform}
	}%
	\subfigure[]{
	\begin{minipage}[t]{0.23\linewidth}
	\centering
	\includegraphics[width=4cm]{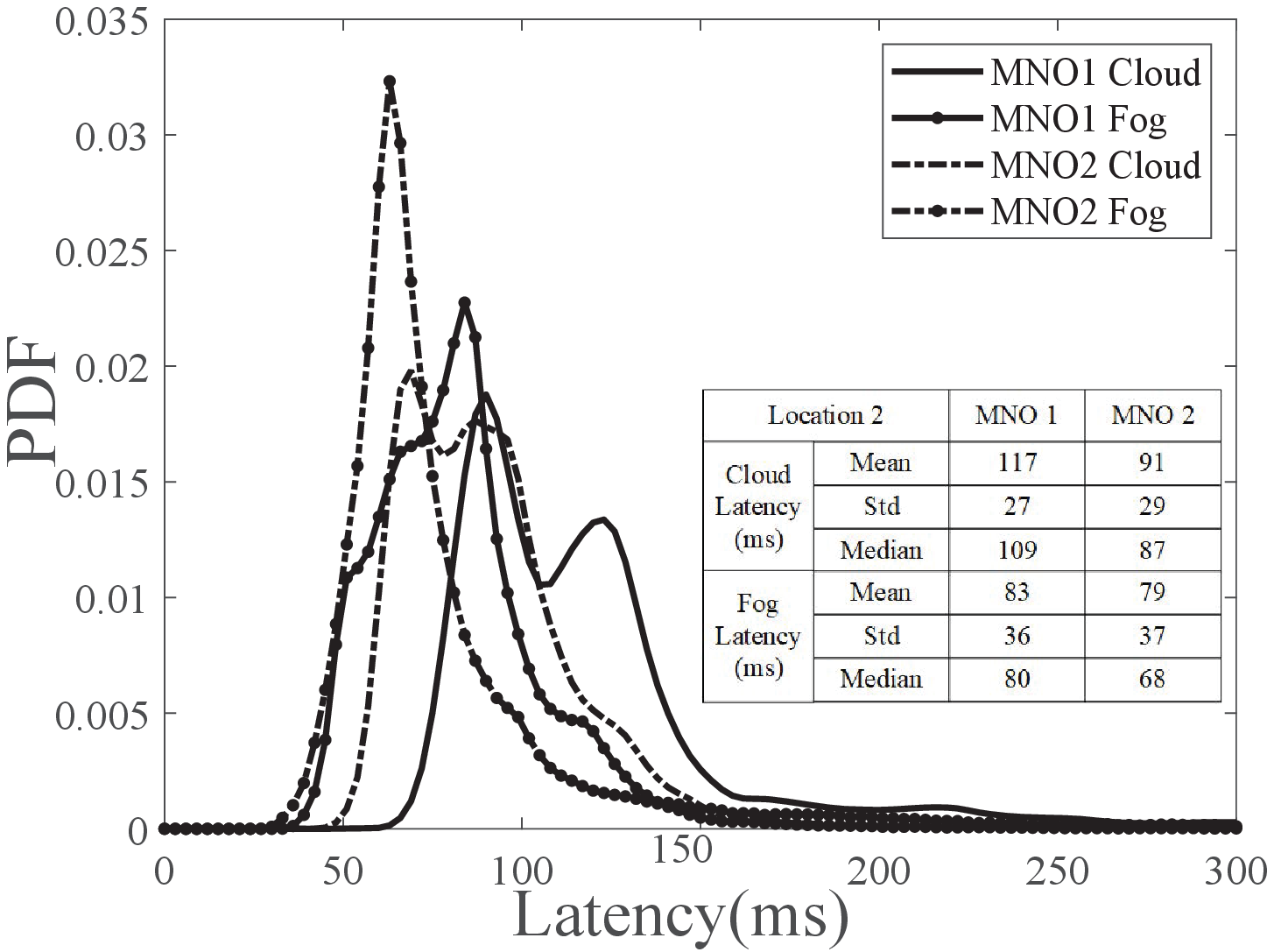}
	\end{minipage}%
	\label{MNO2perform}
	}%
	\subfigure[]{
	\begin{minipage}[t]{0.23\linewidth}
	\centering
	\includegraphics[width=4cm]{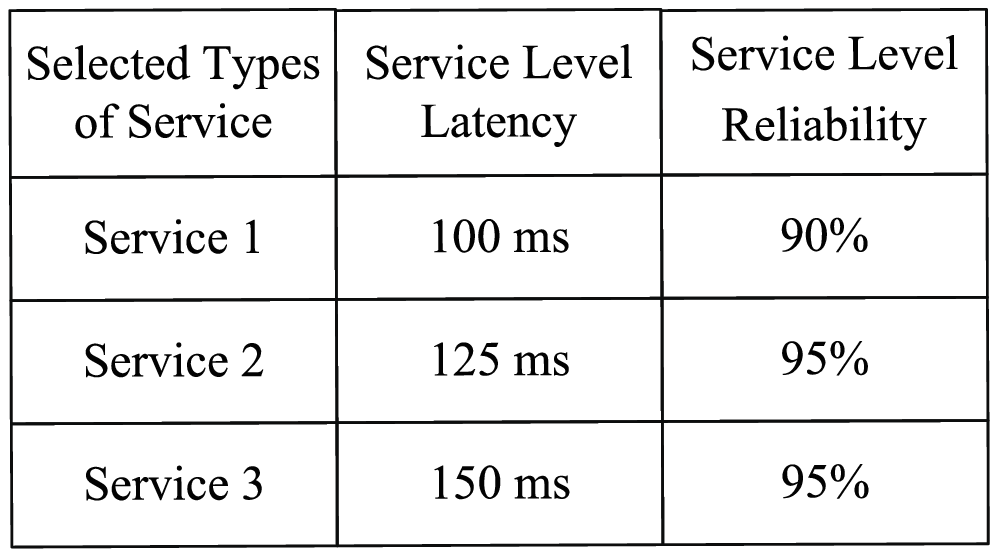}
	\end{minipage}%
	\label{selected_service}
	}
	\subfigure[]{
	\begin{minipage}[t]{0.23\linewidth}
	\centering
	\includegraphics[width=4cm]{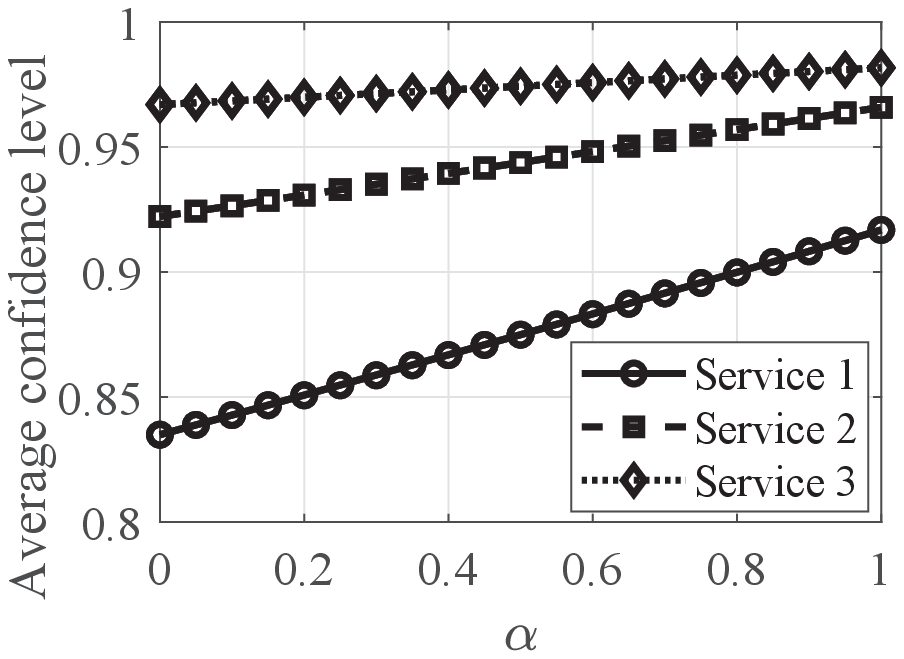}
	\end{minipage}%
	\label{a}
	}
	\subfigure[]{
	\begin{minipage}[t]{0.23\linewidth}
	\centering
	\includegraphics[width=4cm]{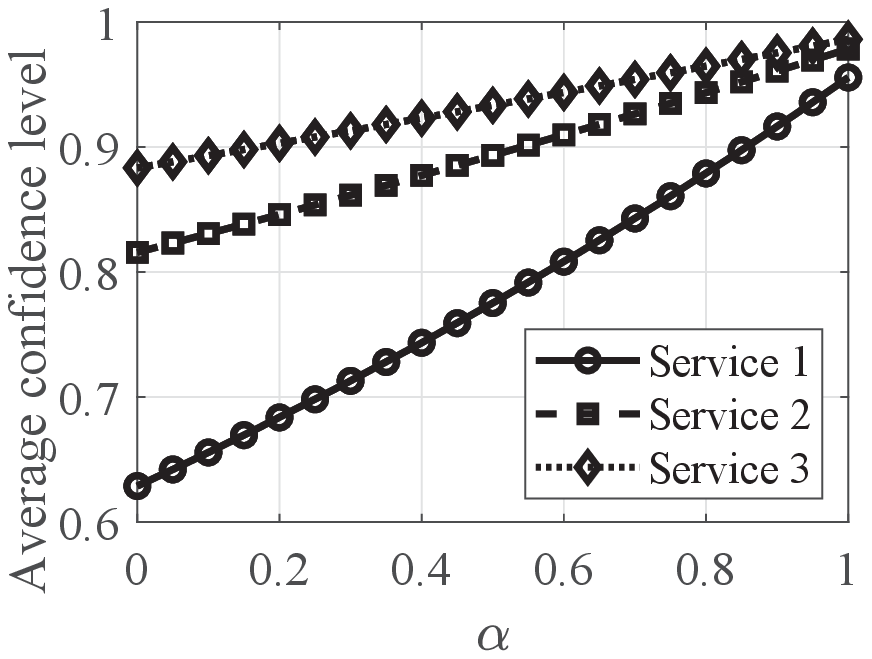}
	\end{minipage}%
	\label{b}
    }
   \subfigure[]{
	\begin{minipage}[t]{0.23\linewidth}
	\centering
	\includegraphics[width=4cm]{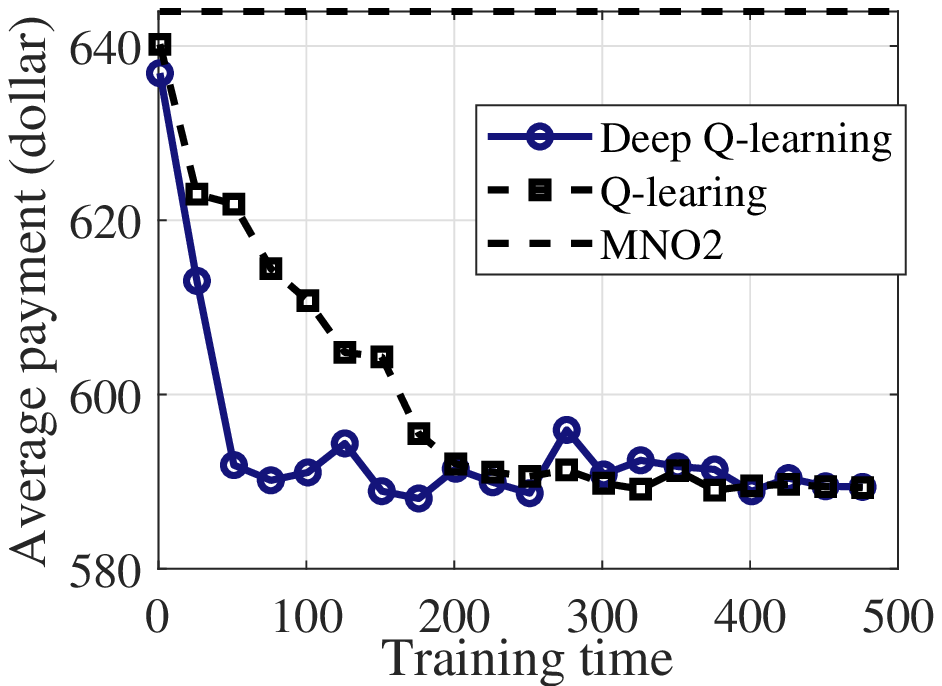}
	\end{minipage}%
	\label{DQN}
	}
    \subfigure[]{
	\begin{minipage}[t]{0.23\linewidth}
	\centering
	\includegraphics[width=4cm]{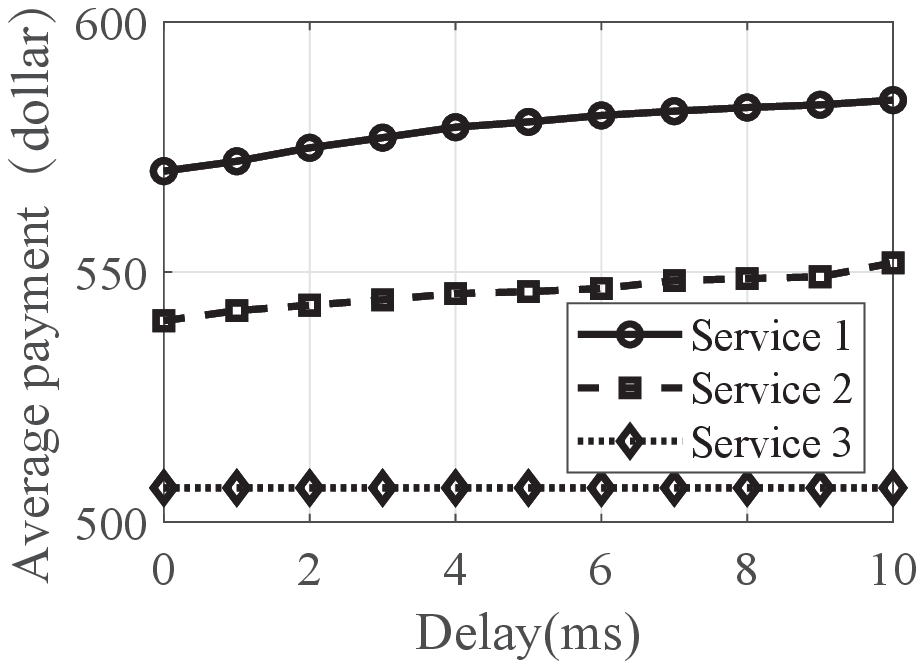}
	\end{minipage}%
	\label{delay}
	}%
	\subfigure[]{
	\begin{minipage}[t]{0.23\linewidth}
	\centering
	\includegraphics[width=4cm]{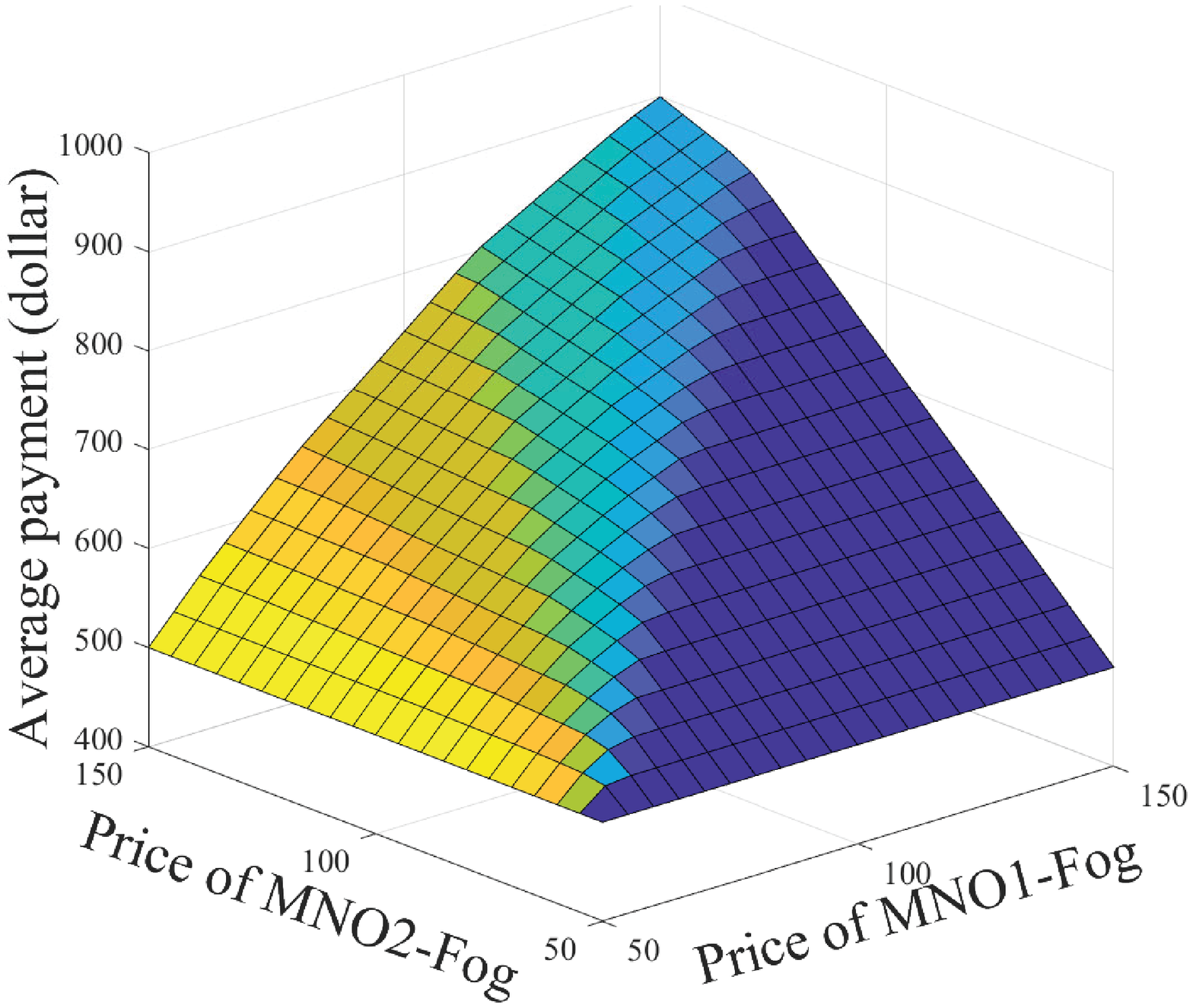}
	\end{minipage}%
	\label{price}
	}
	\label{perform}
	\caption{(a)(b) Empirical PDF of cloud and fog latency of both MNOs at different locations. (c) The requirements of the considered three types of service (d) Average confidence level under different values of $\alpha$ on MNO 1's network. (e) Average confidence level under different values of $\alpha$ on MNO 2's network. (f) Convergence rate of double DQN and Q-learning. (g) Average payments under different values of $d$ with using MNO switching. (h) Average payments under different values of fog price.}
	\end{figure*}
We use the dataset collected from a four-month city-wide measurement. We have developed a smart phone App using Android API to periodically ping the IP address of the most likely fog node location and the Amazon cloud server at every 500 ms. The resulting RTT will be recorded for evaluating the latency performance offered by MNO networks when connecting to cloud and fog servers. 
A smart phone installed with our developed App was mounted on a moving vehicle to collect the RTT data. The RTT data collected in each MNO network consists of two parts: \emph{Cloud latency} and \emph{fog latency}, which are measured by pinging the IP address of the CDC server and the first hop IP address in the core network of each MNO's network, respectively. 
We can then calculate the latency probability distribution of each individual MNO when driving in the main routes of a mid-sized city and use it to evaluate the possible performance of task assignment and MNO switching policy.

In Figures \ref{MNO1perform} and \ref{MNO2perform}, we present the empirical probability distribution of latency when connecting to fog and cloud servers via different MNO networks in two different fixed location measurements. We can observe that the latency performance of MNOs can vary significantly at different time and locations. In particular, MNO 1 provides a better average cloud latency and worse average fog latency performance in location 1, compared to MNO 2. MNO 1 however offers a worse average cloud latency and better fog latency performance in average in location 2.   

To evaluate the impact of the value of $\alpha$, the portion of workload offloaded by fog nodes, on the confidence level of different services, we compare the average confidence level of three types of services measured at a fixed location when connecting to two MNOs' networks. The detailed service requirements are listed in Figure \ref{selected_service}.
We can observe that when increasing the portion of workload to be offloaded by the fog nodes, the increasing rate of the average confidence level offered by MNO 1 as shown in Figures \ref{a} is lower than that achieved by MNO 2. This is because the performance difference between cloud and fog latency in MNO 1 is smaller than that in MNO 2. Also we can observe that for delay-tolerant services, the cloud latency offered by MNO 1 is better than MNO 2. However, when the service is more delay-sensitive, the fog node offered by MNO 2 offers a better performance than that of MNO 1. This again verifies our observations that the service performance of different MNOs can vary significantly.


In Figure \ref{DQN}, we present the convergence rate of double DQN and Q-learning, compared with the average payment when choose a single MNO (MNO 2). 
It can be observed that the proposed double DQN-based approach converges to a neighborhood of the optimal solution within 
100 training time, while Q-learning requires over 200 training time.

In Figure \ref{delay}, we investigate how the extra delay $d$ caused by switching MNOs will influence the utility values of the vehicle. It is obvious that the total cost is a non-decreasing function of the value of $d$. This is because the value of $d$ directly affects the frequency for the vehicle to switch between MNOs, i.e., the higher the value of $d$, the less frequency for the vehicle to change its currently selected MNO. This will reduce the potential performance improvement that can be achieved by MNO switching. Also for delay-tolerance applications (e.g., service 3), it is less sensitive for the change of $d$ because its service requirement can be satisfied by choosing any MNO which reduces the need for MNO switching.  

In Figure \ref{price}, we fix the price of cloud service and compare the average payments made by the considered vehicle under different price charged by fog services offered by two MNOs. We can observe that the total payment is increased with the prices of the fog services. Also the vehicle tends to choose the MNO with the lowest fog price.  

\section{Conclusion}
In this paper, we have studied the potential performance improvement for a cloud/fog computing-supported vehicular system that can be achieved by enabling multi-operator wireless connectivity support. We have formulated the multi-MNO switching problem as a dynamic programming problem taking into consideration the switching cost, service variations of connected vehicle, and different prices charged by cloud and fog servers. To find the optimal MNO switching policy, an optimal workload allocation policy has been introduced. We have then proposed a double DQN method to minimize the cost paid by the considered vehicle with guaranteed latency and reliability. We have evaluated our approach using the dataset collected in a commercially available LTE network for over four months of measurement in a mid-sized city. Numerical results have shown that the double DQN algorithm converges with much shorter training time compared to the traditional Q-learning approach. Also MNO switching policy can significantly reduce the average payment for connected vehicular systems.
\vspace{-0.1in}

\section*{Acknowledgment}
The authors would like to thank Ericsson (China) Hubei Branch and  China Mobile Hubei 5G Joint-innovation Lab for help in the data collection.

\bibliography{ref}
\bibliographystyle{IEEEtran}

\end{document}